# Positron annihilation spectroscopy for heat-treated Fe-Ga alloys and their defect structure

*L.V. El'nikova and Yu.V.Funtikov*

*Institute for Theoretical and Experimental Physics,*
*25 Bolshaya Cheremushkinskaya, Moscow 117218, Russian Federation*

**Abstract**
We present the results of lifetime positron annihilation spectroscopy (PALS) for the ferrum-gallium alloys (galfenols) Fe–21at%Ga and Fe–22.4at%Ga, whose defect structure is connected with annealing from 20 to 1000 ºC along the phase diagram of $Fe_{100}Ga_{100-x}$, at the $L1_2 \rightarrow A_2$ transitions and in the sub-lattice of the $D0_3$ phase. In frames of the standard trapping model, we estimate the positron annihilation parameters for the bulk metal and for the different thermal vacancies herein. Also we define concentrations of these vacancy defects, which may be helpful for explanation of the physical properties of galfenols, including giant softening and extremal magnetostriction at the given intermetallic composition.

**Introduction**

Fe-Ga alloys (galfenols), $Fe_{100}Ga_{100-x}$, possess a remarkable property to change their magnetostriction and elastic softening depending on their composition $x$ [1-8]. At the equilibrium phase diagram of $Fe_{100}Ga_{100-x}$ (Fig. 1), the disordered body-centered cubic (*bcc*) A2 phase (or $\alpha$-Fe) is in equilibrium with an face-centered cubic (*fcc*)–based L12 ordered one. Adding gallium atoms from 17 to 28 at% concentration to the $\alpha$-Fe lattice, one observes, that maximum values of magnetostristion $\lambda_{100}$ up to 400 ppm are known for quenched galfenols with 19 at. % Ga, which is more than an order of magnitude higher than that of pure $\alpha$-Fe [8].

There were a lot of attempts to explain a drastic increase of magnetostricition by specific ordering, accompanied with changes in the type of a defect structure around the Curie point. So, phenomena of linear anelasticity in galfenols are studied with mechanical spectroscopy [1-4], where the relaxation peaks were observed in connection with the presence of linear and point defects, i.e. dislocations, grain and phase boundaries etc. at the crossover regions $A2+D0_3$, $A2+L1_2$, and in the single phases of the equilibrium phase diagram (Fig. 1).

Heat-treatment has a significant influence on structural ordering in galfenols. High resolution X-ray diffraction for alloys Fe–19.5at%Ga and Fe–22at%Ga [2] clarified, that in Fe–19.5at%Ga, after quenching, only the A2 phase exists, whereas, at slow cooling, the 67% A2 and the 33% $D0_3$ phases form. For Fe-22 at% Ga, the two-phase mixture behaves *vice versa*, the 40% A2 and 60% $D0_3$ are observed at quenching and the 100% $D0_3$ appears at slow cooling.

In our consideration, we rest on two compounds Fe–21at%Ga and Fe–22.4at%Ga, where we anticipate appearance of features in defect structure nearby the region of the critical magnetostriction.

At the step-by-step annealing for Fe–21at%Ga, due to the diagram (Fig.1), the regions of the equilibrium phase co-existence ($A2+L1_2$ by 588 ºC are $A2+D0_3$ by 600 ºC) are observed, further with an increase of temperature, the A2 phase extends, whiles for the similar compound Fe–22.4at%Ga, the second crossover comes to 630 ºC. In the A2 phase ($\alpha$-Fe), the Ga atoms are lying in the *bcc* lattice of Fe. But in the range 300-600 ºC, ordering type of the $D0_3$ solution is expected [2]. Following the X-ray data [2, 7, 8], we learnt that at this range, the grain boundaries might be observed at a distortion of the A2 structure. Such linear defects are observable with positron annihilation spectroscopy [9, 10,11] together with point and volume defects as well.

We study Fe–21at%Ga and Fe–22.4at%Ga alloys basing on lifetime positron annihilation spectroscopy, a nondestructive nuclear method, which allows us to probe a material at the atomic level, to distinguish linear, point and volume defects and to describe their density and geometrical adjectives depending of physical-chemical conditions [12,13].

The single known data on PALS for Fe–19at%Ga [5] are not enough to characterize the total defect structure at annealing for the galfenols compositions of industrial interests. The present work serves to revealing defects structure in another similar thermal treated specimens and helps in explanation of their physical-, in particular, magnetostrictive properties to include these views on galfenols in a common picture of PAS mechanisms in iron composites at the atomic level.

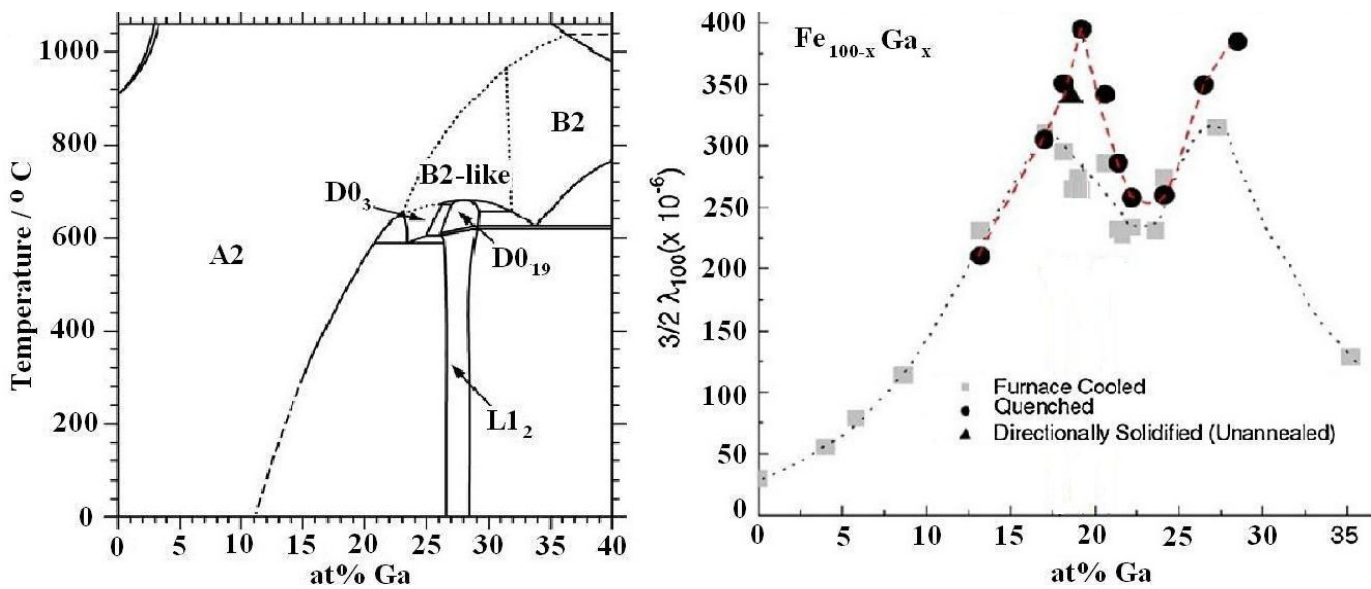

Fig.1. Equilibrium phase diagram of $Fe_{100}Ga_{100-x}$ alloys (left) and their magnetostriction $\lambda_{100}$ [2] (right).

**Experimental**

In experiments on the positron spectrometer «Time» (Fig.2), the positron lifetimes were measured, $e^+$ from radioactive source come from aqueous $^{22}$NaCl (20 µCi activity) with the energy 511 keV in the Lavsan 5µm thickness foil arranged together with two surround specimens of sizes 12×18×4 mm$^2$ in the "sandwich-type" geometry. With the software Resolutionfit – PALSfit [14], its contribution in spectra is found to be 20.3%. The spectrometer resolution is of two components (so called FWHM, "full-width-at-half-maximum") $FWHM_1$ =247 ps and $FWHM_2$ =490 ps, the graduating mark of the spectrometer is 6 ps. The lifetime measurements were carried out due to the standard fast-coincidence method. The number of events in a peak per a channel was $10^6$.

Initially the specimens were quenched for 1000 °C. Annealing of Fe–21at%Ga was in the range 20, 150…1000 °C, they were in cooling at every step from 150, 200, 250, 300, 350, 400, 450, 500, 600, and 1000 °C. At 600°C, there was quenching, and at other temperatures, cooling has been performed in a furnace during two hours. Analogously for the Fe–22.4at%Ga specimens, the annealing steps were 150, 250, 350, 450, 550, 625 (for further quenching), and 1000 °C. Annealing with quenching have to fix the D0$_3$ phase in the A2 matrix, and annealing with furnace cooling have to fix the L1 phase in A2 respectively.

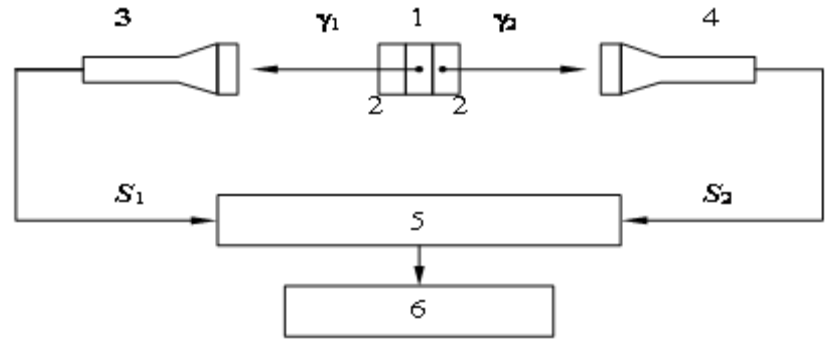

Fig. 2. Block-scheme of the PALS experiment. 1 is the positron source of activity (1-2) ×10⁶ Bq, 2 is the test specimen; two photomultipliers 3,4 to record the START and STOP γ-quanta convert light signals into digital ones $S_1$ and $S_2$, which are processed in the electronic module 5 to be passed to a computer 6 [13].

**Results and discussions**

In the PALS spectra for galfenols Fe–21at%Ga and Fe–22.4at%Ga, the positron lifetimes $\tau_1, \tau_2, ...$ and corresponding intensities $I_i, I_2, ...$ may be distinguished.

In terms of the standard trapping model [15], for our galfenols, lifetime spectra may be well decomposed into two exponential components; from which, one may extract positron annihilation in the bulk of the galfenol and annihilation in voids. The bulk lifetime $\tau_b$ is calculated due to the formulae $\tau_b = 1/\lambda_b$, where the annihilation rate $\lambda_b$ is expressed as follows:

$$\lambda_b = I_1 \tau_1^{-1} + I_2 \tau_2^{-1}. \qquad (1)$$

Here we apply for spectra processing the two-state trapping model, where one of states is associated with linear and point defects (mono-, divacancy *etc*.), and second one is of their clusters. However, we neglect vacation clusters as unobservable in such systems at annealing.

Also to characterize the defect distribution in galfenols in the annealing steps, we consider the trapping rate $k_d$ putting inverse time, which positron needs to find a void to annihilate therein:

$$k_d = \frac{I_2}{I_1}\left(\frac{1}{\tau_b} - \frac{1}{\tau_2}\right). \qquad (2)$$

For Fe–21at%Ga and Fe–22.4at%Ga, the received average positron lifetime $\tau_{av}$, $\tau_{av} = \Sigma(\tau_i I_i)/\Sigma I_i$, the lifetimes $\tau_{bulk}$, $\tau_1$ and $\tau_2$ and their intensities are shown in Figs 3-6.

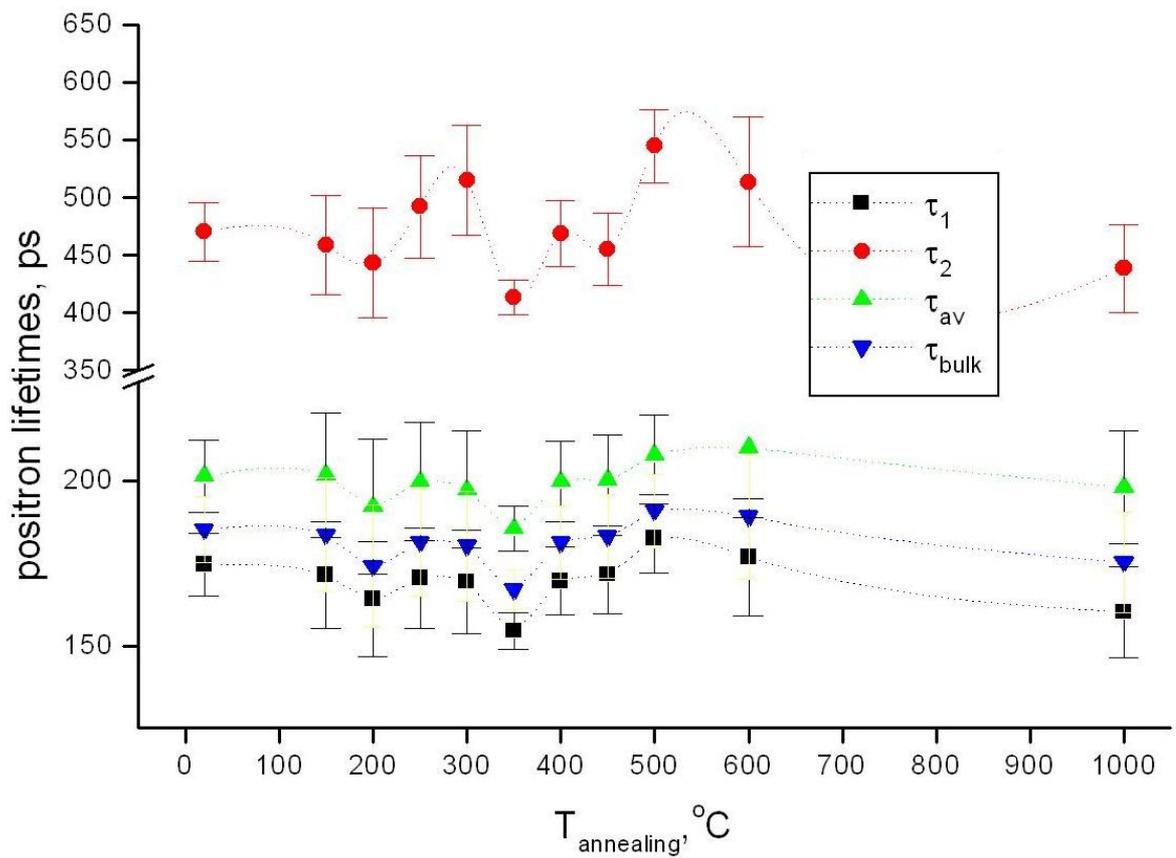

Fig. 3. Measures positron lifetimes $\tau_1$ and $\tau_2$ and calculated $\tau_{av}$ and $\tau_{bulk}$ vs annealing temperatures in Fe–21at%Ga

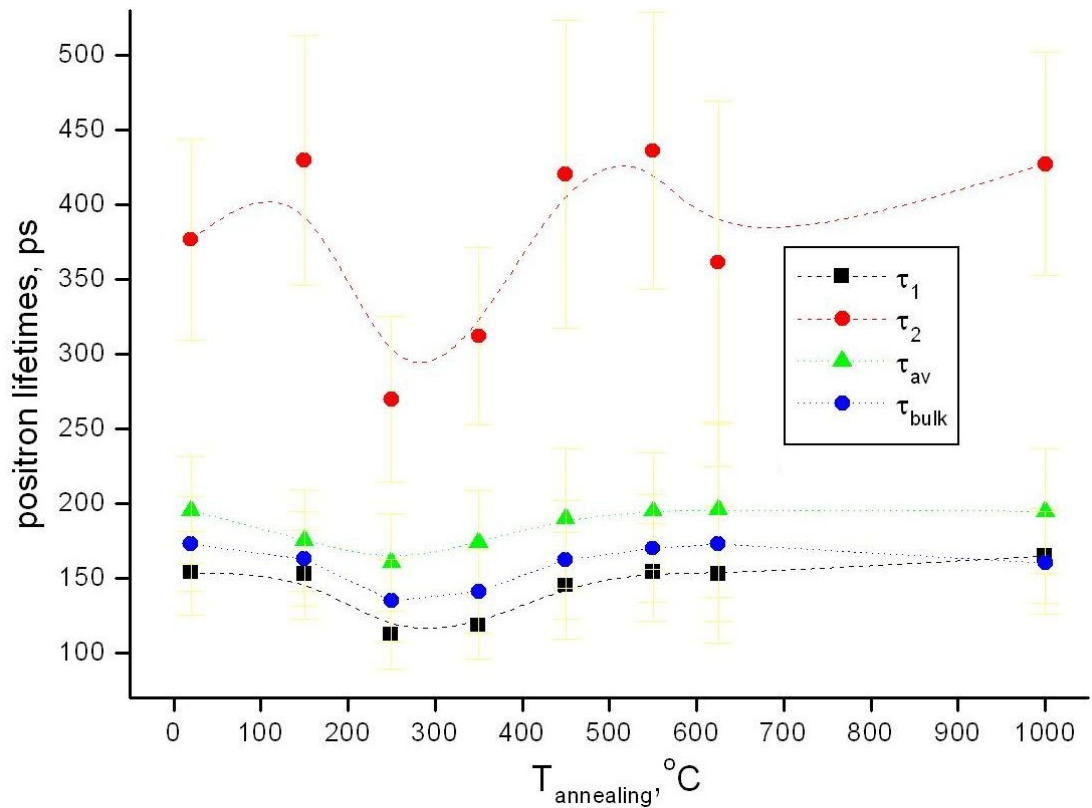

Fig. 4. Measures positron lifetimes $\tau_1$ and $\tau_2$ and calculated $\tau_{av}$ and $\tau_{bulk}$ vs annealing temperatures in Fe–22.4at%Ga

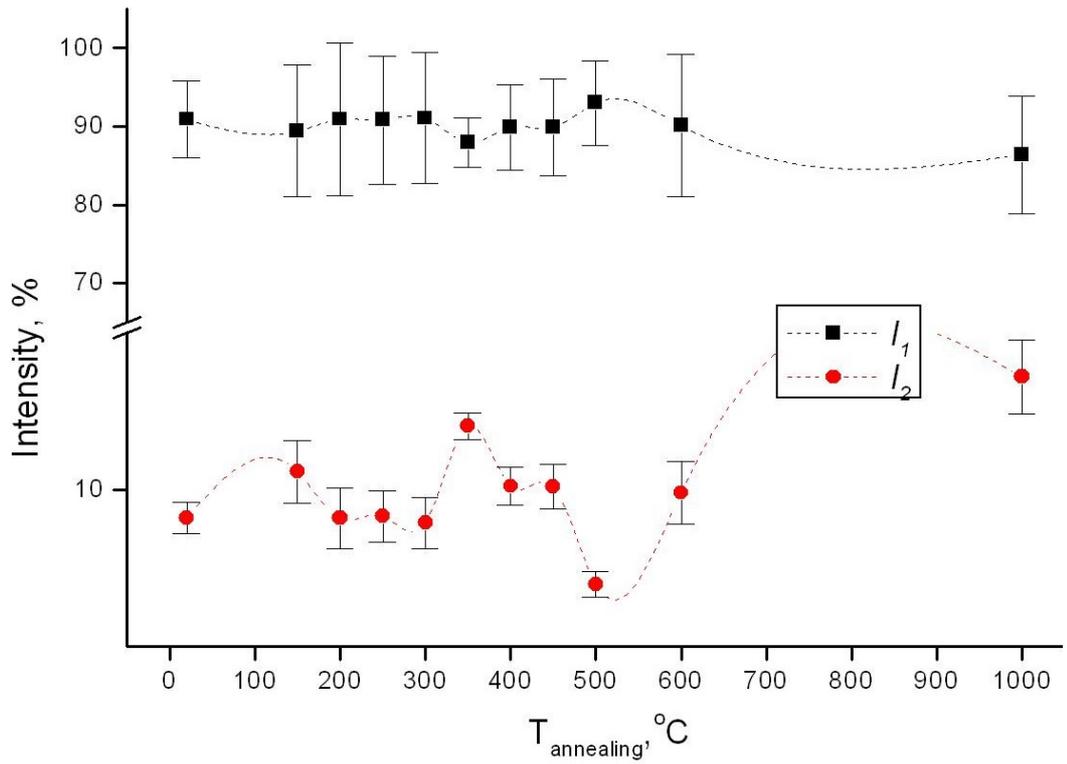

Fig. 5. Intensities of positron lifetime components $\tau_1$ and $\tau_2$ vs annealing temperature in Fe–21at%Ga

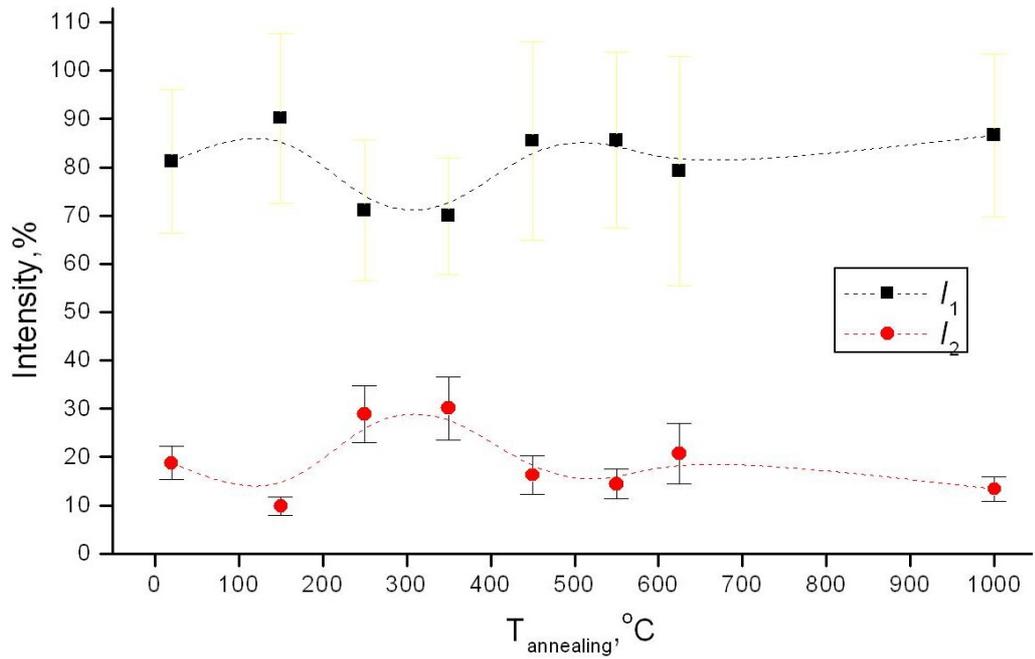

Fig. 6. Intensities of positron lifetime components $\tau_1$ and $\tau_2$ vs annealing temperature in Fe–22.4at%Ga

Table 1. Components of PALS spectra in Fe–21at%Ga

|  | 20 °C | 150 °C | 200 °C | 250 °C | 300 °C | 350 °C | 400 °C | 450 °C | 500 °C | 600 °C | 1000 °C |
|---|---|---|---|---|---|---|---|---|---|---|---|
| $\tau_1$, ps | 174.6 ±9.4 | 171.5 ±16.1 | 164.1 ±17.6 | 170.5 ±15.3 | 169.4 ±15.6 | 154.4 ±5.6 | 169.6 ±10.3 | 171.6 ± 11.8 | 182.6 ±10.6 | 176.92 ±17.9 | 160.2 ±14.0 |
| $I_1$,% | 90.9 ±4.9 | 89.5 ±8.4 | 91.0 ± 9.7 | 90.9 ± 8.2 | 91.1 ± 8.4 | 88.0 ± 3.2 | 89.9 ± 5.5 | 89.9 ± 6.2 | 93.0 ± 5.4 | 90.1 ±9.1 | 86.4 ±7.5 |
| $\tau_2$, ps | 470.7 ±25.4 | 458.7 ±43.1 | 443.4 ±47.4 | 492.3 ±44.3 | 515.2 ±47.4 | 413.7 ±14.9 | 469.0 ±28.6 | 455.6 ±31.4 | 545.0 ±31.6 | 513.62 ±56.5 | 438.6 ±38.2 |

| | | | | | | | | | | |
|---|---|---|---|---|---|---|---|---|---|---|
| $I_2$,% | 9.1±0.5 | 10.6±1.0 | 9.1±1.0 | 9.2±0.8 | 8.9±0.8 | 12.0±0.4 | 10.1±0.6 | 10.1±0.7 | 7.0±0.4 | 9.9±1.0 | 13.6±1.2 |

Table 2. Components of PALS spectra in Fe–22.4at%Ga

| | 20 °C | 150 °C | 250 °C | 350 °C | 450 °C | 550 °C | 625 °C | 1000 °C |
|---|---|---|---|---|---|---|---|---|
| $\tau_1$, ps | 153.40±20.5 | 152.3±29.7 | 112.0±23.0 | 112.1±22.1 | 144.8±25.4 | 154.0±23.3 | 152.4±45.7 | 147.6±31.7 |
| $I_1$, % | 81.20±11.0 | 90.15±17.6 | 71.08±14.6 | 67.17±12.1 | 83.70±14.6 | 85.55±13.1 | 79.22±23.7 | 86.62±16.9 |
| $\tau_2$, ps | 376.7±70.5 | 429.6±83.8 | 269.8±55.3 | 300.7±59.2 | 420.5±68.4 | 436.1±63.1 | 361.3±108.4 | 348.6±75.0 |
| $I_2$,% | 18.80±7.3 | 9.85±1.9 | 28.92±5.9 | 32.83±6.5 | 16.30±6.8 | 14.45±8.7 | 20.78±6.2 | 13.38±2.6 |

Registered positron lifetimes (Fig.3) correspond to positron annihilation in a bulk alloy and in defect regions, e.g. in dislocations or grain boundaries, mono-vacancies, divacancies and vacation clusters. It is known, that in $\alpha$-Fe, the positron lifetimes in dislocations are 165 ps, and in mono-vacancies, divacancies and 3-,4-,6-vacation clusters they are 175, 197, 232, 262, 305 ps respectively, and corresponding sizes of defects are as follows: dislocations have a crosscut radius 1.3Å [14], and mono-, divacancies and 3-,4-,6-vacation clusters have radii R = 1.41, 1.6, 1.9, 2.2 and 2.6Å respectively [16].

The specific trapping constant for the defects is

$$\chi_d = 4\pi \cdot D \cdot R = \sigma \cdot v. \quad (3)$$

It depends on size (radii) of a defect $R$ and on the positron diffusion coefficient $D$. $\sigma$ is a positron capture cross-section ($\sigma = \pi R^2$), $v$ is the thermal velocity of positron, $v \approx 1.05 \cdot 10^7$ cm/s at room temperature.

The positron thermal velocity depends on temperature due to the following expression

$$v = \sqrt{\frac{8k_B T}{\pi m_+^*}}, \quad (4)$$

where $k_B$ is the Boltzmann constant, and mass of quasi-free positron is $m_+^* \approx 9.1 \cdot 10^{-28}$ g.

From (3), (4), we found the defect concentrations $c_d = k_d/\chi_d$ [11]. In table 3 for reader's cogency and in Figs.7-9, we show the positron thermal velocity $v$ and the defect concentration $c_d$ at different annealing steps for Fe–21at%Ga and For Fe–22.4at%Ga.

Table 3. Positron thermal velocity and concentrations of different type defects for Fe–21at%Ga

| $T$,°C | 20 | 150 | 200 | 250 | 300 | 350 | 400 | 450 | 500 | 600 | 1000 |
|---|---|---|---|---|---|---|---|---|---|---|---|
| $v$, $10^7$cm/s | 1.050 | 1.262 | 1.270 | 1.336 | 1.400 | 1.458 | 1.516 | 1.570 | 1.624 | 1.726 | 2.084 |
| $c_d$,$10^{15}$ m$^{-2}$ dislocations | 1.512 | 1.492 | 1.390 | 1.348 | 1.315 | 1.585 | 1.304 | 1.246 | 0.959 | 1.170 | 1.334 |
| $c_d$,$10^{17}$ cm$^{-3}$ monovacancies $R$=1.4 Å | 0.533 | 0.522 | 0.469 | 0.448 | 0.432 | 0.571 | 0.428 | 0.380 | 0.256 | 0.345 | 0.420 |
| $c_d$,$10^{17}$ cm$^{-3}$ divacancies $R$=1.6Å | 0.388 | 0.381 | 0.342 | 0.327 | 0.315 | 0.417 | 0.312 | 0.277 | 0.196 | 0.264 | 0.322 |
| $c_d$,$10^{17}$ cm$^{-3}$ 3-vacancies $R$=1.6Å | 0.275 | 0.270 | 0.243 | 0.232 | 0.223 | 0.295 | 0.222 | 0.206 | 0.139 | 0.188 | 0.228 |

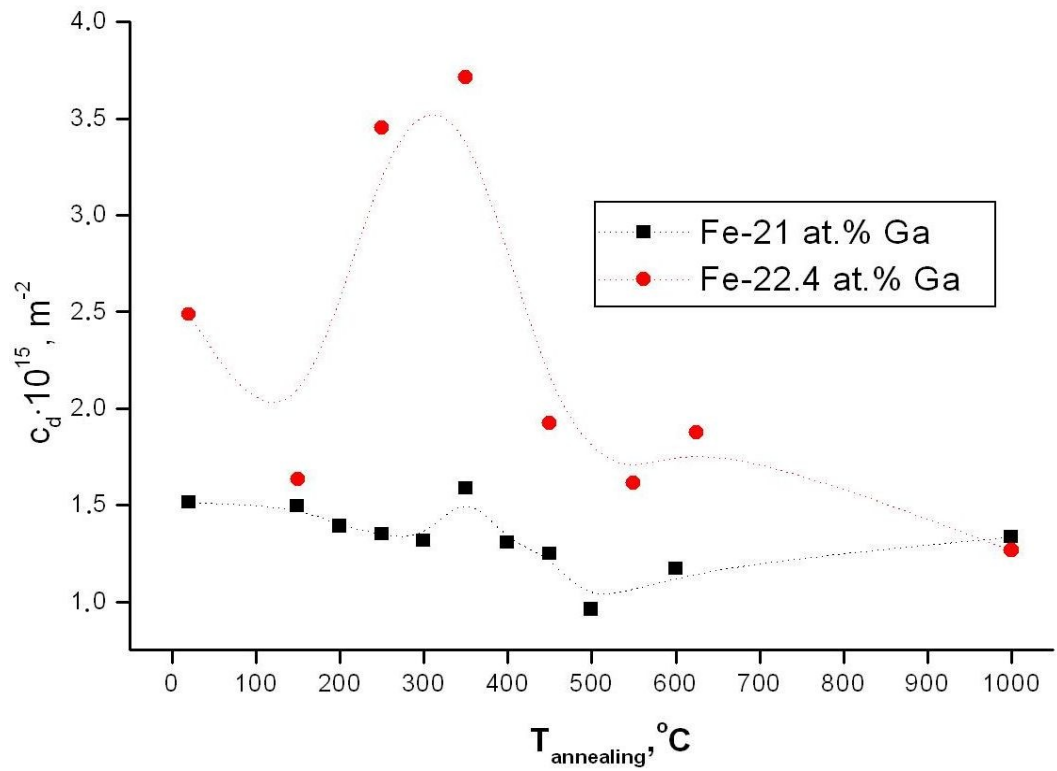

Fig. 7. Concentrations of linear defects *vs* annealing temperature in Fe–21at%Ga and Fe–22.4at%Ga

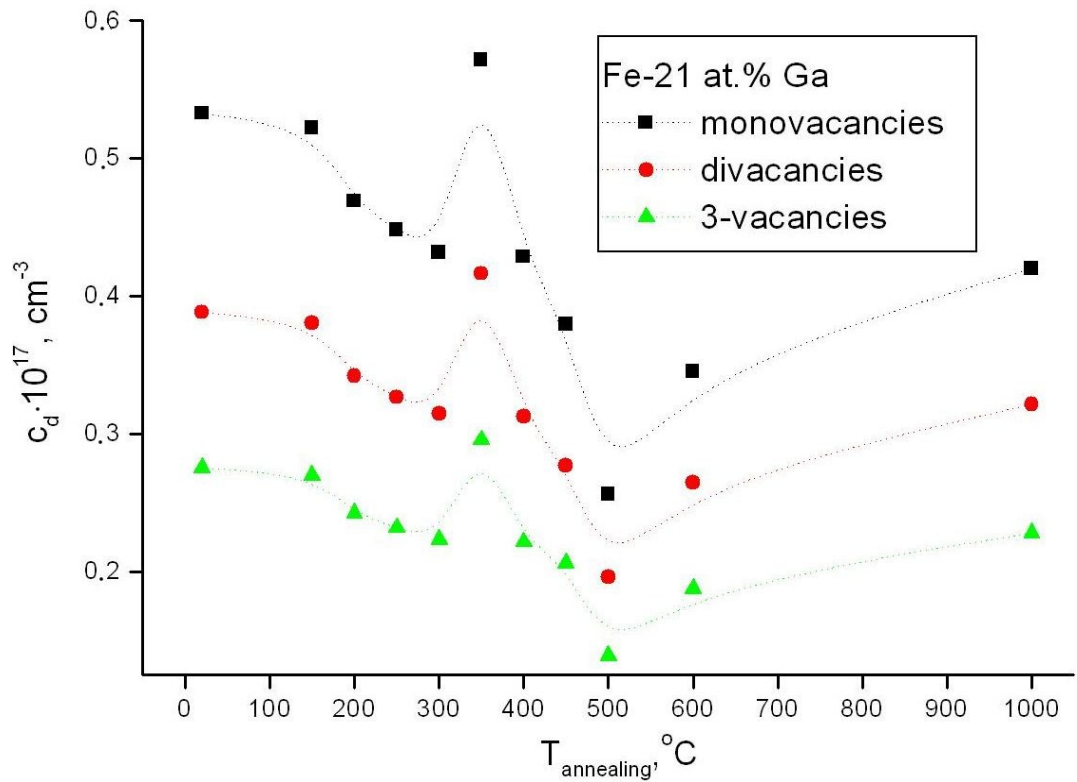

Fig. 8. Concentrations of volume voids *vs* annealing temperature in Fe–21at%Ga

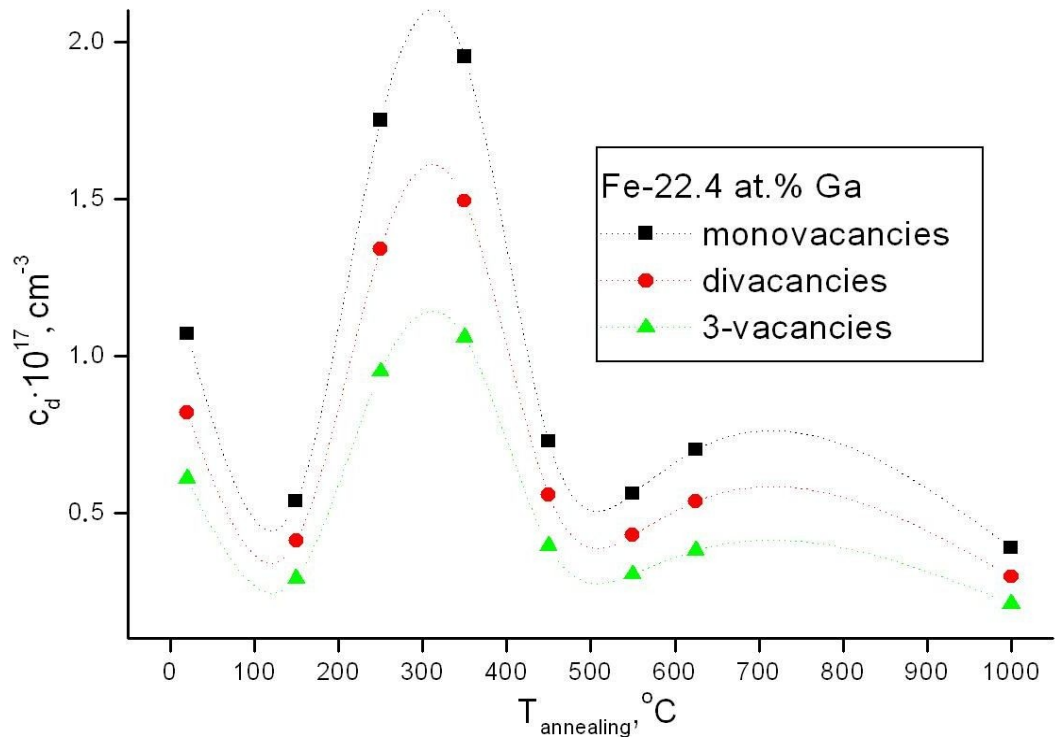

Fig. 9. Concentrations of volume voids *vs* annealing temperature in Fe–22.4at%Ga

In behavior of the defect density, we reveal uniformity for both alloys. The defect concentration decreases at annealing by 150 °C, typically for metals [15]. A few differences are observed in linear defects (dislocations) comparing with point (volume) ones.

At an increase of temperature, dislocations became almost annealed at 300 °C Fe–21at%Ga and at 150 °C in Fe–22.4at%Ga, but appear again up to 350°C, next, they vanish up to 500 °C, and appear slightly at increase of further annealing. At 1000 °C, the dislocation density is equal $1.3·10^{15}$ m$^{-2}$ for both galfenols (Fig. 7).

Volume defects have similar distributions in Fe–21at%Ga and Fe–22.4at%Ga (Fig. 8,9) over the annealing steps, but positions and depth of their minima up to 350 °C is different (at 300 and 150 °C respectively). It means, that Fe–22.4at%Ga has weaker magnetoctrictive properties, than Fe–21at%Ga due to the phase diagram Fig.1.

The void concentration has to correlate with hardness and magnetostriction [5, 8] in alloys and in galfenols especially. To explain the singularity at 350 °C in both galfenols, we use the data on the maximal shear modulus for the case of the structural-isomeric galfenol Fe–17at%Ga [5]. In addition, for the alloy Fe–18at%Ga, this point 350 °C may confirm the hypothesis on the ordering of the $DO_3$ bcc sub-lattice in the A2 fcc phase [1].

**Conclusions**

1. The bulk lifetime $τ_{bulk}$ in Fe–21at%Ga and Fe–22.4at%Ga galfenols is found to be greater, than that well-known 110 ps in pure iron.
2. A typical single-metal behavior for Fe–21at%Ga and Fe–22.4at%Ga galfenols hold at the range 20-150 °C.

3. High ordering phases with minimal $c_d$ for volume defects are observed at annealing temperature 500 °C in compliance with the phase diagram Fig.1 (at the phase transition from A2+L1$_2$ to A2+D0$_3$ at 588 °C). But lowest minimum of $c_d$ at 300 °C and maximal $c_d$ at 350 °C lies in the same region of coexisting phases.
4. At the annealing temperature 350 °C corresponding to highest magnetostriction, linear and volume defects present in maximal values.
5. To study the bulk structures of these galfenols, the PALS method should be supplemented with other techniques, e.g X-ray diffraction, mechanical spectroscopy, hardness testing *etc*. However, to characterize only defects in a material, it is a sufficient probe.

**Acknowledgments**